\documentclass[aps,prl,twocolumn,showpacs,superscriptaddress,floatfix,nofootinbib]{revtex4-1}
\usepackage{amsfonts,hyperref}

\usepackage{graphicx}
\usepackage{amsmath}
\usepackage{amssymb}

\newcommand{\hc}{\hat c}

\begin{document}
\title{Competing states in the $t$-$J$ model: uniform d-wave state versus stripe state}

\author{Philippe Corboz}
\affiliation{Theoretische Physik, ETH Zurich, 8093 Zurich, Switzerland}
\affiliation{Institute for Theoretical Physics, University of Amsterdam,
Science Park 904 Postbus 94485, 1090 GL Amsterdam, Netherlands}

\author{T. M. Rice}
\affiliation{Theoretische Physik, ETH Zurich, 8093 Zurich, Switzerland}

\author{Matthias Troyer}
\affiliation{Theoretische Physik, ETH Zurich, 8093 Zurich, Switzerland}

\date{\today}

\begin{abstract}
Variational studies of the $t$-$J$ model on the square lattice based on infinite projected-entangled pair states (iPEPS) confirm an extremely close competition between a uniform d-wave superconducting state and different stripe states. The site-centered stripe with an \emph{in-phase} d-wave order has an equal or only slightly lower energy than the  stripe with \emph{anti-phase} d-wave order. The optimal stripe filling is not constant but increases with $J/t$. A nematic anisotropy reduces the pairing amplitude and the energies of stripe phases are lowered relative to the uniform state with increasing nematicity. 
\end{abstract}

\pacs{71.10.Fd, 71.27.+a, 71.10.Hf}

\maketitle

The discovery of high-temperature superconductivity in the cuprates stimulated intense study of the $t$-$J$ model~\cite{zhang88} - the strong coupling limit of the Hubbard model~\cite{Hubbard63,Anderson87}, on a square lattice. But open issues remain concerning the phase diagram at underdoping, especially with regard to the stability and form of stripe phases. Initially these were considered to be simple charge- and spin-density waves with enhanced hole doping along $\pi$-domain walls in an antiferromagnetic (AF) background at a filling of one hole per unit length per stripe ~\cite{poilblanc89,zaanen89,machida89,schulz89}. Later theoretical work found that half-filled stripes with coexisting d-wave superconducting (SC) order~\cite{white98_tJ,white99_competition}, or even more complex order with intertwined domain walls in both the AF and d-wave SC order~\cite{himeda02}   are very close competitors to states  with uniform hole density. The proposal by Berg et al ~\cite{berg07} that the latter stripe form explained the observation by Li et al~\cite{li07} of 2-dimensional superconductivity order over a large temperature range in La$_{2-x}$Ba$_x$CuO$_4$  around $x=1/8$ stimulated further theoretical investigations. Surprisingly many calculations on the $t$-$J$ model using a range of different approximations found  small energy differences between states with uniform hole density and the stripe states~\cite{himeda02,raczkowski07,chou08,yang09,chou10}. This near degeneracy between states with clearly different ordering suggests an underlying general physical explanation. This interpretation is further supported by the experimental observation of the stripe state in a specific hole density range in some cuprates. (See Refs.~\cite{emery99,kivelson03,birgeneau06,ogata08,vojta09,fradkin10,fradkin12} for a review). 

In this paper we use an improved version of the powerful infinite projected entangled-pair states (iPEPS) method on the $t$-$J$ model. This method yields the lowest energy variational wavefunctions to date for infinite (or very large) two-dimensional systems. It gives remarkably small energy differences for the very different stripe and uniform states. Interestingly, as the accuracy of the method is increased, the energy differences between the competing states become \emph{smaller}. Our version of the $t$-$J$ model ignores the usual next-nearest neighbor hopping for computational simplicity, but this omission did not affect the near degeneracies in the earlier calculations~\cite{himeda02,chou08,yang09}, suggesting an underlying general physical explanation, which remains to be uncovered. 

The near degeneracy, on the one hand, makes the identification of the true ground state extremely difficult, on the other hand it implies that the  $t$-$J$ model in the physically relevant regime is at or close to a phase transition between competing phases.  So small additional and/or anisotropic terms in the model can stabilize one phase over the other. Since these additional terms will depend on the particular cuprate compound, it can explain why stripes are only found in certain cuprates.  As an example of a modified $t$-$J$ model, we study the effect of a nematic anisotropy, which can be introduced by the tilting pattern of the CuO$_6$ octahedra e.g. in  the low-temperature tetragonal (LTT) phase of La$_{2-x}$Ba$_x $CuO$_4$  around $x=1/8$,   and confirm that it lowers the energy of the stripe state relative to the uniform state.

\emph{Model --}
The $t$-$J$ model is given by the Hamiltonian
\begin{equation}
\hat H= - t \sum_{\langle ij \rangle \sigma} \left( \tilde{c}_{i \sigma}^{\dagger}\tilde{c}_{j\sigma}  + H.c.\right) +  J\sum_{\langle ij \rangle}  \left( \hat S_i \hat S_j - \frac{1}{4} \hat n_i \hat n_j\right)  
\end{equation}
with ${\langle ij \rangle}$  nearest-neighbor pairs, $\sigma=\{\uparrow,\downarrow\}$ the spin index, $\hat n_i=\sum_\sigma \hc^\dagger_{i \sigma} \hc_{i \sigma}$ the electron density and $\hat S_i$ the spin $1/2$ operator on site $i$, and $\tilde{c}_{i\sigma}=\hc_{i\sigma} ( 1 - \hc^\dagger_{i \bar \sigma} \hc_{i \bar \sigma})$. 

\emph{Method  --} \
Our results are obtained with (fermionic) infinite projected-entangled pair states (iPEPS) - a variational tensor network ansatz to efficiently represent two-dimensional ground states in the thermodynamic limit~\cite{verstraete2004,Verstraete08,jordan2008,corboz2010}.
 It can be seen as a natural generalization of matrix product states (the underlying ansatz of the density-matrix renormalization group method~\cite{white1992}) to two dimensions.  Originally it has been developed for spin systems, and later extended to fermionic systems~\cite{Corboz10_fmera, kraus2010, pineda2010,  barthel2009, shi2009, Corboz09_fmera,corboz2010,pizorn2010,gu2010}. 
   The ansatz consists of a supercell of rank-5 tensors which is periodically repeated on the lattice. 
     Each tensor has a physical index and four auxiliary indices which connect to the nearest-neighboring tensors. The accuracy of the ansatz can be systematically controlled by the bond dimension $D$ of the auxiliary indices (each tensor contains $3D^4$ variational parameters).
  A $D=1$ iPEPS simply corresponds to a site-factorized wave function (product state), and by increasing $D$ quantum fluctuations (or entanglement) can be systematically added to the state.  
 A similar ansatz has been employed in Ref.~\cite{corboz2011}, however, here we use a more accurate optimization scheme (the full update, cf. Ref.~\cite{corboz2010}) to find the best variational parameters. We also push the simulations to larger bond dimensions by exploiting U(1) symmetries~\cite{singh2010,bauer2011} and  a more efficient contraction method (see supplemental material~\cite{SM}). 

\begin{figure}[]
\begin{center}
\includegraphics[width=1\columnwidth]{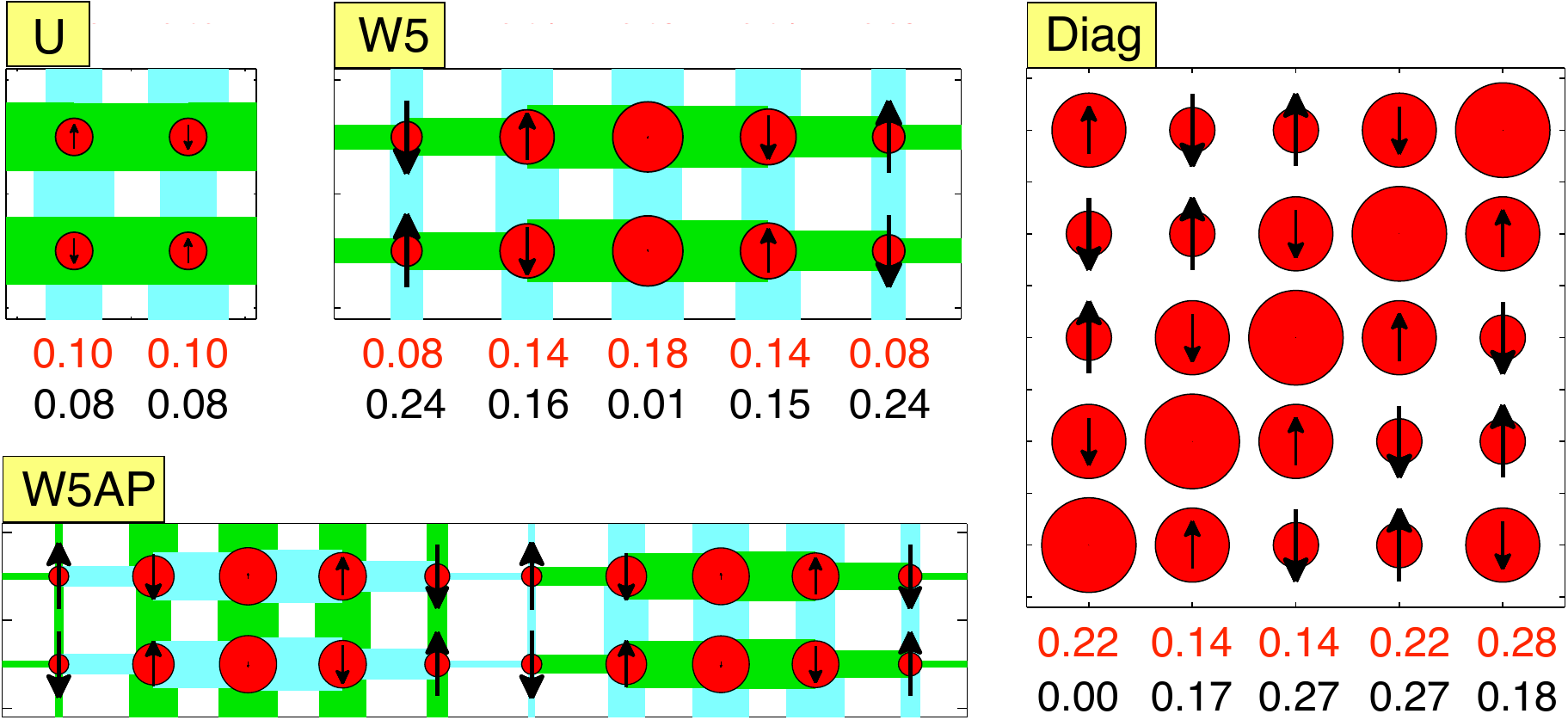} 
\caption{(Color online) Competing low-energy states in the $t$-$J$ model found with iPEPS simulations using different supercells ($J/t=0.4$).
The diameter of the red dots (length of the arrows) is proportional to the local hole density (local magnetic moment) with average values given by the first (second) row of numbers below a panel. The width of a bond between two sites scales with the (singlet) pairing amplitude on the corresponding bond with different sign in horizontal and vertical direction indicated by the two different colors. 
[U]~Uniform d-wave superconducting state with coexisting antiferromagnetic order ($\delta \sim 0.1$, $D=14$), where two different tensors for the two sublattices have been used. [W5]~A site-centered vertical stripe state of width $W=5$ with in-phase d-wave order in a $5\times2$ supercell  ($\delta\sim 1/8$, $D=14$). [W5AP]~A site-centered stripe state of width $W=5$ with anti-phase d-wave order in a $10\times 2$ supercell  ($\delta\sim1/8$, $D=10$). [Diag]~A fully-doped ($\rho_l=1$), insulating diagonal stripe in a $L\times L$ cell using $L$ different tensors at a doping $\delta=1/L$ (here $L=5$, $D=14$). We considered sizes up to $L=11$. }
\label{fig:states}
\end{center}
\end{figure}

We compare various competing low-energy states in the $t$-$J$ model by using different supercell sizes in iPEPS, e.g. a uniform state with d-wave SC  order coexisting with AF order at low doping, and different types of stripe states, with examples presented in Fig.~\ref{fig:states}. 
Each panel shows several order parameters computed with iPEPS: the hole density $\delta_i = 1- \langle \hat n_i \rangle$ and the local magnetic moment $\hat S^z_i$ on each site $i$, and the singlet pairing amplitude  $\Delta = \langle \hat c_{i \uparrow} \hat c_{j \downarrow} - c_{j \downarrow} \hat c_{i \uparrow} \rangle / \sqrt{2} $ between neighboring sites $i$ and~$j$.



\begin{figure}[]
\begin{center}
\includegraphics[width=1\columnwidth]{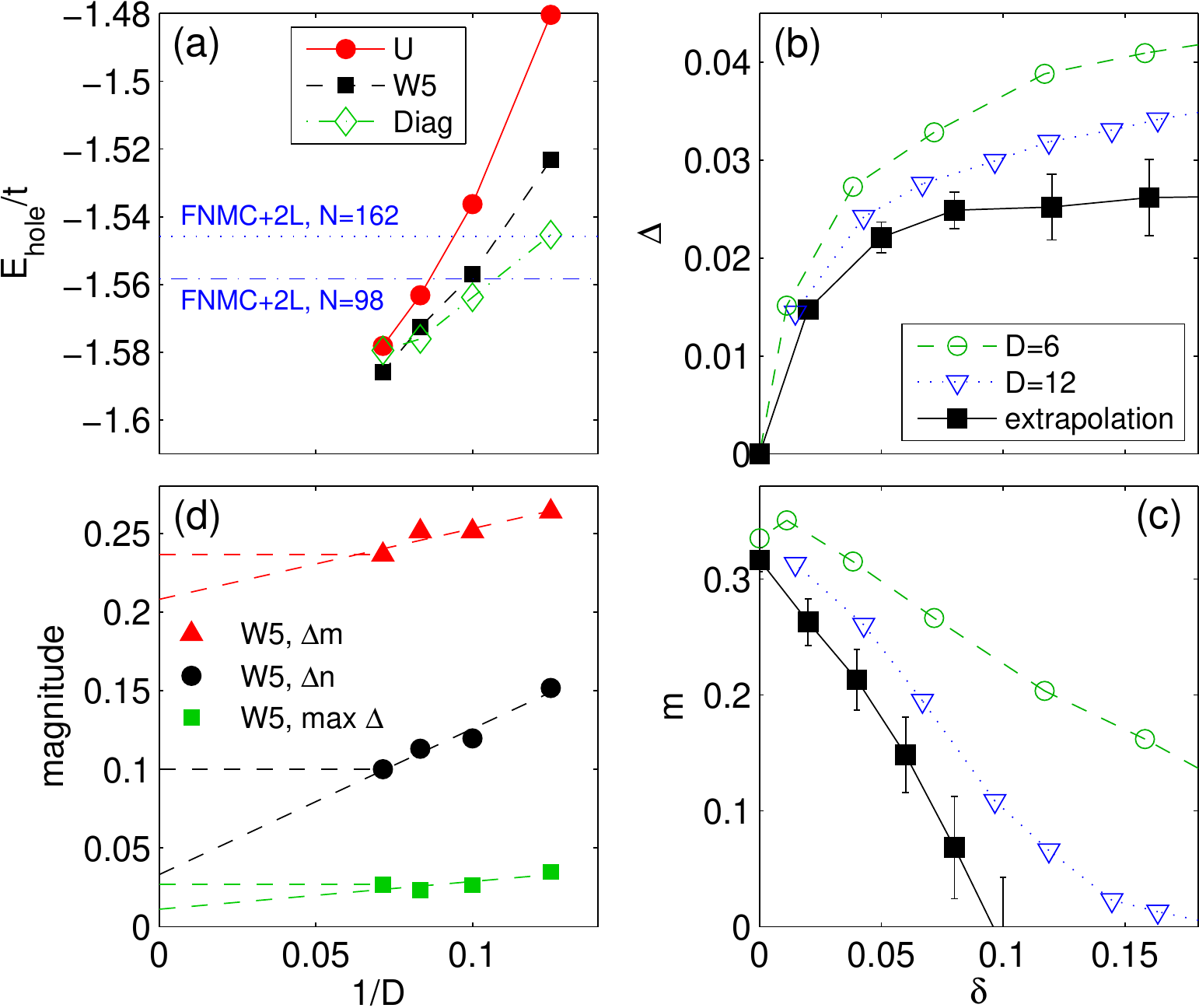}
\caption{(Color online) (a) Energies of the competing states as a function of inverse bond dimension for $\delta= 0.12$. The horizontal lines show the best fixed-node Monte Carlo result (with 2 Lanczos steps) from Ref.~\cite{hu12}. (b)-(c) Order parameters of the uniform d-wave state as a function of doping: (b)~the pairing amplitude $\Delta$ and (c) the local magnetic moment $m$. The extrapolated values have been obtained from a linear extrapolation of the finite $D$ data, which provides a rough estimate of the order parameters in the infinite $D$ limit. 
(d)  Order parameters of the W5 stripe state as a function of inverse $D$ for $\delta= 0.12$: the modulation strength of the local hole density $\Delta n=n_{\max} - n_{\min}$ and of the local magnetic moment $\Delta m=m_{\max} - m_{\min}$, where $n=\langle \hat n \rangle$ and $m=|\langle \hat S_z \rangle |$ are evaluated on each lattice site in the supercell. The filled squares show the maximal singlet pairing $|\Delta|$. The order parameters decrease with increasing $D$, but remain finite in the infinite $D$ limit. The dashed lines are a guide to the eye. 
}
\label{fig:res_ntJ04apaper7}
\end{center}
\end{figure}

\emph{Uniform d-wave state --} 
We first discuss the results obtained with an iPEPS consisting of only two tensors, one for each sublattice, for $J/t=0.4$. The lowest energy state we find with this ansatz has a uniform charge distribution and a d-wave SC order,  coexisting with AF order at low doping (see [U] in Fig.~\ref{fig:states}). A similar state has been found in several previous studies~\cite{dagotto93,himeda02,ogata03,ivanov04,shih04,lugas06, spanu08,hu12}, however, here we obtain a  lower variational energy for this state than the best result from fixed-node Monte Carlo combined with two Lanczos steps (FNMC+2L)~\cite{hu12}, see Fig.~\ref{fig:res_ntJ04apaper7}(a). For example, at doping $\delta=0.12$ we find an energy per hole $E_{hole}=(E_s - E_0)/\delta =  -1.578t$ for $D=14$, where $E_s$ is  the energy per site and $E_0=-0.467775$ the value at zero doping taken from Ref.~\cite{sandvik1997}. This value is considerably lower than $E_{hole}=-1.546t$ obtained for a system with $N=162$ in Ref.~\cite{hu12}, where the  energy increases with system size.

In Fig.~\ref{fig:res_ntJ04apaper7}(b) we present results for the singlet pairing amplitude $\Delta$ of the uniform state as a function of doping, for $D=6$, $D=12$ and the extrapolated data in $1/D$ (see \cite{SM} for additional data). It is suppressed with increasing $D$, but tends to a finite value in the infinite $D$ limit,  $\Delta\approx 0.025$ for $\delta=0.12$. The local magnetic moment $m$ shown in Fig.~\ref{fig:res_ntJ04apaper7}(c) decreases rapidly with doping, and is also suppressed with increasing $D$. For $\delta\lesssim0.1$ the extrapolated value of $m$ in $1/D$ is finite, but it vanishes for larger $\delta$.  Thus, we find coexisting d-wave and antiferromagnetic order for $\delta\lesssim0.1$ in close agreement with previous results~\cite{himeda02,ogata03,ivanov04,shih04,lugas06, spanu08,hu12}.

\emph{Stripe states --}
Next we  focus on vertical stripe states, which are obtained with supercells of size $P\times2$ with $P$ the periodicity of the stripe. Each stripe has a certain width $W$ given by the periodicity of the charge density wave order (which is not necessarily equal to $P$), and a filling measured in holes per unit length of a stripe, $\rho_l = W \delta $.  In Refs.~\cite{white98_tJ,corboz2011} it was found that the preferred width of a stripe increases with decreasing doping (see also~\cite{SM}), i.e. depending on the doping we need to use different supercell sizes. To simplify the discussion we will focus on $W=5$ stripes in the following which in our calculations are energetically favored for dopings around $\delta\sim0.12$~\cite{commentW4}.

The lowest energy $W=5$ stripe we find is the W5 state shown in Fig.~\ref{fig:states}.
This state exhibits a modulation in the charge-, spin-, and superconducting order, where the maximal doping is centered on a row of sites, called site-centered stripe (as opposed to bond-centered stripes, see Refs.~\cite{vojta08,vojta09, greiter10,greiter11} for a discussion). 
Fig.~\ref{fig:res_ntJ04apaper7}(d) shows that both the amplitudes of the charge- and spin- modulation decrease with increasing bond dimension $D$, but then tend to a finite value in the infinite $D$ limit, which indicates that in this state the stripe order persists in this limit.

The d-wave pairing in the W5 stripe state has the same sign structure on neighboring stripes, i.e. in-phase order. In agreement with previous  studies~\cite{himeda02,raczkowski07,chou08,yang09,white09_tJ,scalapino12} we also find a competing low-energy state which has anti-phase order (W5AP in a $10\times2$ supercell shown in Fig.~\ref{fig:states})  with an energy per hole that is only slightly higher (of the order of $0.001t$ for $D=10$) than the in-phase stripe, see \cite{SM} for additional data. 
 Since the energy difference between the two states is very small, it is conceivable that anti-phase stripes get stabilized by additional terms (such as a next-nearest neighbor hopping~\cite{himeda02}). This further supports the proposal that anti-phase ordered stripes are the reason for the lack of 3D superconductivity above $T=4$K in La$_{2-x}$Ba$_{x}$CuO$_4$ around $x=1/8$~\cite{li07,huecker11}, because they lead to a suppression of the interlayer Josephson coupling between the copper-oxygen planes~\cite{berg07}.

Finally, we also find diagonal stripes with a low energy, e.g. the state shown in the right panel in Fig.~\ref{fig:states}. These states are obtained by using supercells of size $L\times L$ with $L$ different tensors arranged in a diagonal stripe pattern.  These stripes are insulating and have a filling of $\rho_l=1$ holes per unit length. However, we will show in the next section that diagonal stripes are energetically unfavorable at large $D$.

\emph{Uniform vs stripe states --}
So far, we have found various low energy states in different supercells. Next we make a systematic comparison of  their energies for $J/t=0.4$ and $\delta=0.12$, to determine which of the competing states is the true ground state. 
For a fixed value of $D=8$  we find that the uniform state has a higher variational energy than the W5 stripe state, in agreement with previous findings~\cite{corboz2011}. Furthermore, it turns out that diagonal, insulating stripes  - which were not considered in Ref.~\cite{corboz2011} -  are even lower in energy for $D=8$. 
 However, from this we cannot conclude that the diagonal stripe state is the ground state, but we must examine \emph{how the energies of the competing states change upon increasing $D$}, shown in Fig.~\ref{fig:res_ntJ04apaper7}(a): All energies decrease with increasing $D$, however with different slopes, such that the W5 stripe state becomes lower in energy than the diagonal stripe state for $D>12$. For $D=14$ the W5 stripe state has the lowest energy, but since the energy of the uniform state decreases  faster (at least for $D<12$) than the energy of the W5 state it may get lower (or equal) in the large $D$ limit. Such a crossing of energies of competing states as a function of $D$ has already been found in another model~\cite{Corboz13_su3hc} and it is a possible scenario also for the present case. 

Even if we cannot conclusively determine the ground state based on our results, the important message from our data is that the uniform and the vertical stripe state are still strongly competing at considerably lower variational energies than in previous studies for large 2D systems~\cite{hu12}. It thus seems likely that both states play an important role for the low-energy physics of the $t$-$J$ model, and that small perturbations (e.g disorder, open boundaries~\cite{hellberg99}, etc.) in the system can be enough to stabilize different states. However, our data shows that diagonal stripes are energetically higher than vertical stripes. [We have not found evidence for the stable diagonal stripes  observed in experiments~\cite{wakimoto99,birgeneau06} in the low doping limit in the present model.]

\emph{Remarks on phase separation --} While it is well established that the $t$-$J$ model undergoes phase separation for large $J/t$ and small doping~\cite{emery90,putikka92,valenti92,kohno97,shih98,calandra98},  some previous studies predicted phase separation to occur also in the physically relevant regime $J/t\sim0.4$ (see e.g.\cite{hellberg97,ivanov04}). In our study we do not find evidence for phase separation, at least not in the doping regime $\delta \gtrsim 0.08$ (see \cite{SM} for a discussion).

\begin{figure}[tb]
\begin{center}
\includegraphics[width=1\columnwidth]{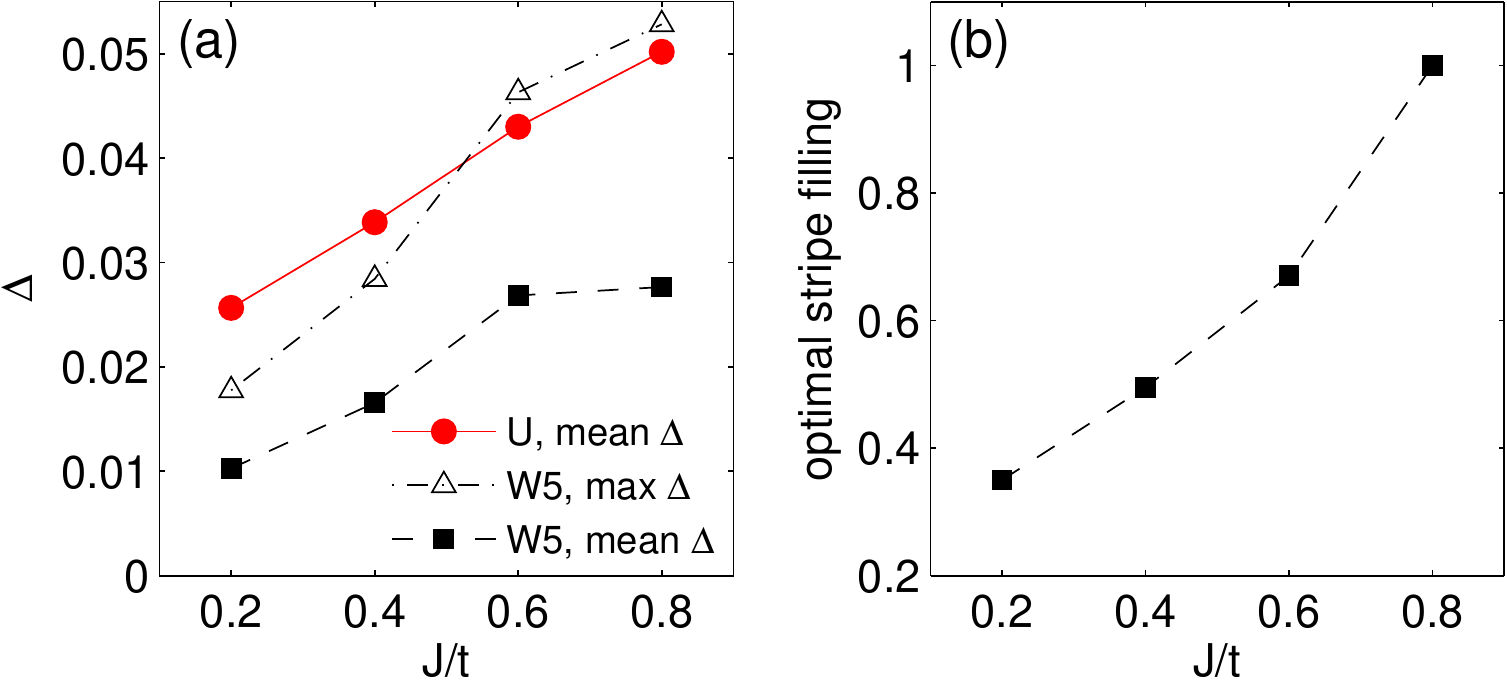}
\caption{(Color online) (a)~Pairing amplitude as a function of $J/t$ for $\delta=0.14$, $D=10$. (b)~Optimal stripe filling $\rho_l=W \delta$ as a function of $J/t$ ($D=10$, $W=5$).}
\label{fig:ntJ02}
\end{center}
\end{figure}

\emph{Other values of $J/t$ --}
It is conceivable that the close competition between the uniform and the vertical stripe state may be a specific feature for $J/t=0.4$. This motivated us to do a similar study also for other values of $J/t$ to check if we can detect a clear phase transition between the two states as a function of $J/t$. However, for small values $J/t=0.2$ as well as for large values $J/t=0.8$ we find  a qualitatively similar dependence on $D$ as in the $J/t=0.4$ case, i.e. the uniform state is higher than the stripe state, but they become closer and closer with increasing $D$. Thus, the strong competition between the two states can be found for a wide range of $J/t$. We also computed the pairing amplitude as a function of $J/t$, shown in Fig.~\ref{fig:ntJ02}(a) for $\delta=0.14$, which increases with $J/t$ for both states, with almost a linear dependence for the uniform state.

A rather unexpected finding concerns the optimal stripe filling, i.e. the filling at which the energy per hole has a minimum for a stripe of a fixed width.  Several previous studies predicted that the minimum is at $\rho_l=0.5$ holes per unit length (i.e. half-filled stripes)~\cite{white98_tJ,white98_energetics,white99_competition,chou10,corboz2011} , which is in close agreement with our results for $J/t=0.4$. However,  here we find that this is  only true for $J/t \sim 0.4$, and that the optimal stripe filling actually depends continuously on $J/t$, i.e. it is a function of the physical parameters of the system. Fig.~\ref{fig:ntJ02}(b) shows that  for  $J/t=0.2$ the optimal $\rho_l$ is $\approx 0.35$, i.e. smaller than half filling, whereas for $J/t=0.8$ the minimum energy per hole is found for a fully-doped stripe ($\rho_l=1$). 

\begin{figure}[b]
\begin{center}
\includegraphics[width=1\columnwidth]{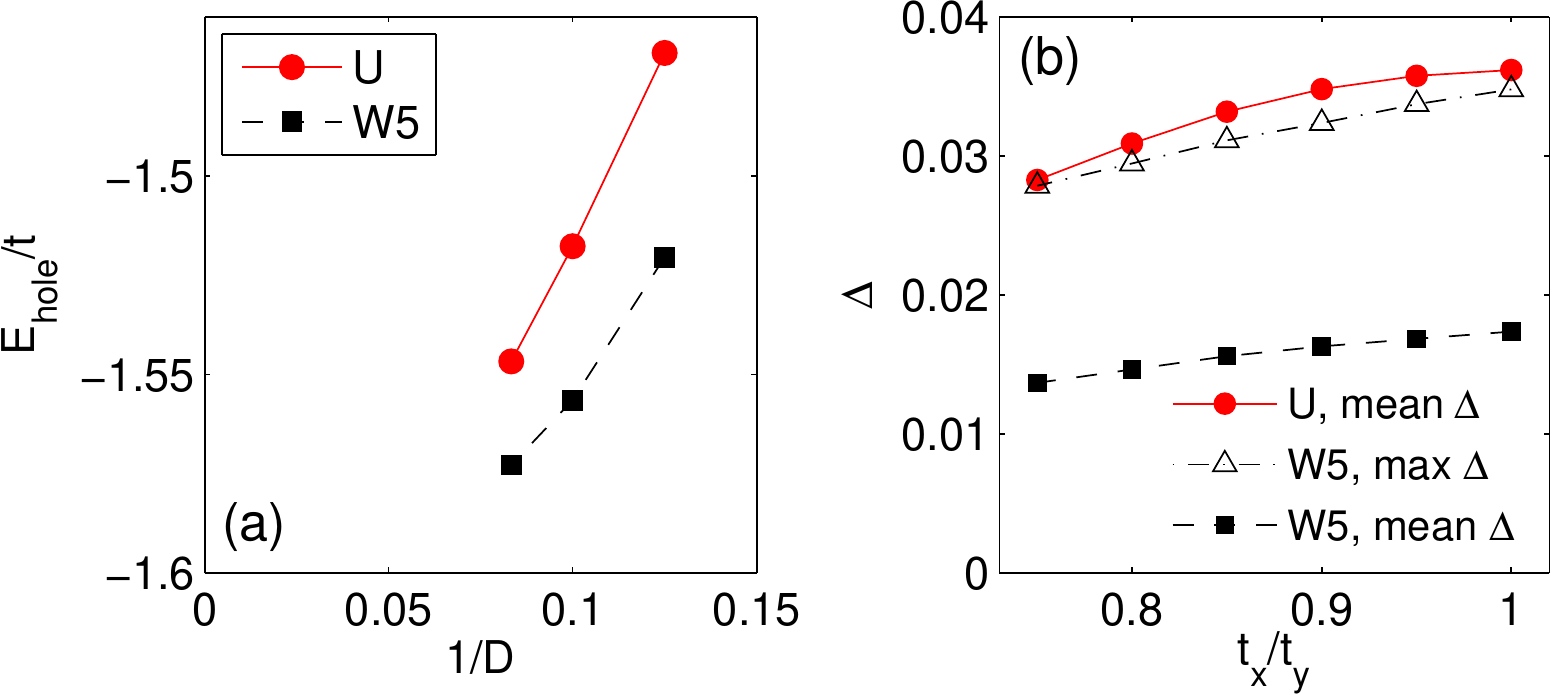}
\caption{(Color online) (a) Energies as a function of inverse bond dimension for $\delta=0.1$ in the nematic $t$-$J$ model with $J/t=0.4$ and $t_x=0.85 t$. (b) The pairing amplitude of the uniform state decreases with increasing nematicity (here  $(t_x+t_y)/2=1$ and $(J_x/t_x + J_y/t_y)/2=0.4$ are kept constant, $\delta=0.13$, $D=8$). }
\label{fig:aniso}
\end{center}
\end{figure}

\emph{Nematic case --}
Motivated by the fourfold rotational lattice symmetry breaking in each CuO$_2$ layer in the LTT phase of La$_{2-x}$Ba$_x$CuO$_4$ and related compounds around $x=1/8$ we study the effect of a nematic anisotropy in the $t$-$J$ model. In Fig.~\ref{fig:aniso}(a) we show the results for $t_x=0.85 t_y$ and $J_x=(0.85)^2 J_y$ with $J_y/t_y=0.4$, at a doping $\delta=0.1$. 
 Comparing with the isotropic case, the vertical W5 stripe state has lowered its energy with respect to the uniform state, which shows that nematicity helps to stabilize the stripe state, in agreement with previous findings~\cite{normand01,becca01,capello08}. We also find that the optimal stripe filling is shifted towards smaller doping, around $\rho_l\approx 0.4$, see~\cite{SM}.

At low doping the preferred orientation of the stripe is along the direction with stronger couplings, i.e the y-direction in this case as found in Refs.~\cite{becca01,kampf01,capello08} (and in Ref.~\cite{normand01} for non-superconducting stripes). However, we find that at large doping ($\delta\gtrsim0.14$) it is the opposite orientation which is preferred, i.e. horizontal stripes. This can be understood by looking at the energy contributions in the two spacial directions in the isotropic case (see \cite{SM} for the individual energy contributions): For a vertical stripe around half filling the exchange term $E^J$ is dominant over the kinetic term $E^\text{kin}$, and it is stronger (lower) in y- than in x-direction, $E^\text{kin}_y<E^\text{kin}_x$. Thus, in the nematic case the stripe can minimize its energy by orienting itself \emph{parallel} to the direction with stronger couplings. However, for large $\delta$ it is the transverse kinetic energy $E^\text{kin}_x$ which is dominant, since with increasing doping $E^J$ becomes weaker. Furthermore, $E^J_x < E^J_y$ at large doping, so that for the total energy we find $E^\text{tot}_x < E^\text{tot}_y$. Thus, in the nematic case at large doping it is favorable for the stripe to form perpendicular to the direction with stronger couplings. [A similar conclusion for fully-doped stripes has been reached in Ref.~\cite{normand01}.]

Finally, we study the effect of the nematicity on the pairing amplitude, shown in Fig.~\ref{fig:aniso}(b). For both the uniform and the stripe state we find that the pairing amplitude is suppressed  with increasing nematicity, i.e. the maximal pairing is obtained in the isotropic case.

\emph{Conclusion --}
Even with a substantially higher accuracy than in previous studies, and in the limit of an infinite system where boundary and finite size effects are negligible, we still find an extremely close competition between the uniform and the vertical stripe state. The origin of this near degeneracy remains a crucial open question and requires further theoretical investigation. One possibility is that  the nearest-neighbor $t$-$J$ model is at or close to a phase transition which separates the two states, i.e. small additional terms in the Hamiltonian can be enough to stabilize one of the states. These additional terms  depend on the particular cuprate compound, and we believe that studying the effect of these terms will explain why stripes appear in certain materials whereas other compounds show no signs of stripes. For example, here we confirmed that a nematic anisotropy, which can be found in the LTT phase of La$_{2-x}$Ba$_x$CuO$_4$, favors the stripe state over the uniform state.

We have studied the properties of the competing states individually: 
the uniform state has d-wave order coexisting with antiferromagnetic order for $\delta\lesssim 0.1$. The pairing amplitude increases with $J/t$ approximately linearly and gets suppressed with increasing nematicity. The vertical stripe state is site-centered and has a finite modulation amplitude of the spin and charge order. Stripes with anti-phase  order have a similar or only slightly higher energy than stripes with in-phase order. 
In the presence of a nematic anisotropy the stripe orientation depends on the doping. Finally, we have shown that the optimal stripe filling is not necessarily $\rho_l=0.5$, but depends on $J/t$. Therefore, a theory of the physics of stripes should include the optimal stripe filling as a free parameter.

\section{Acknowledgments}
We thank W.-J. Hu, F. Becca, and S. Sorella for providing us their data from Ref.~\cite{hu12} and S. Kivelson and E. Fradkin for many useful comments. The support from the Swiss National Science Foundation is acknowledged. The  simulations have been performed on the Brutus cluster at ETH Zurich. This work is part of the D-ITP consortium, a program of the Netherlands Organisation for Scientific Research (NWO) that is funded by the Dutch Ministry of Education, Culture and Science (OCW).

\bibliographystyle{apsrev4-1}
\bibliography{../bib/refs,tJ2comments}

\begin{thebibliography}{75}
\bibitem [{com()}]{commentW4}%
  \BibitemOpen
  \href@noop {} {}\bibinfo {note} {In experiments $W=4$ stripes have been
  found, which in our calculations have a slightly higher energy than $W=5$
  stripes, see \cite{SM}. However, this may change for $t'\neq 0$.}\BibitemShut
  {Stop}%
\end{thebibliography}

\end{document}


\title{Competing states in the $t$-$J$ model: uniform d-wave state versus stripe state: \textit{supplemental material}}

\author{Philippe Corboz}
\affiliation{Theoretische Physik, ETH Zurich, 8093 Zurich, Switzerland}
\affiliation{Institute for Theoretical Physics, University of Amsterdam,
Science Park 904 Postbus 94485, 1090 GL Amsterdam, The Netherlands}

\author{T. M. Rice}
\affiliation{Theoretische Physik, ETH Zurich, 8093 Zurich, Switzerland}

\author{Matthias Troyer}
\affiliation{Theoretische Physik, ETH Zurich, 8093 Zurich, Switzerland}

\date{\today}

\maketitle

\section{Comments on the iPEPS method}

An infinite projected-entangled pair state (iPEPS)~\cite{jordan2008,corboz2010} is an efficient ansatz for two-dimensional ground states in the thermodynamic limit. It is made of a rectangular supercell of size $L_x \times L_y = N_T$, containing $N_T$ rank-5 tensors, $A^{[x,y]}$ labelled by the position $[x,y]$ relative to the supercell~\cite{corboz2011}. This supercell is periodically repeated on the infinite lattice.  If the wave function is translational invariant, a supercell with only one tensor can be used (i.e. the same tensor is repeated on each lattice site). However, since we work directly in the thermodynamic limit, translational symmetries (and other symmetries of the model) may be spontaneously broken. In this case a larger super cell which is compatible with the ground-state structure is required. 
For the uniform state described in the main text we take 2 different tensors arranged in a checkerboard order (in order to include the possibility of antiferromagnetic long-range order). The W5 stripe state and the W5AP state require 10 and 20 different tensors, respectively. The diagonal stripe states in the $L \times L$ cells are obtained by using $L$ different tensors, repeated in the cell compatible with the stripe pattern.

The optimization of the tensors is done via an imaginary time evolution  using a second order Trotter-Suzuki decomposition. For the involved truncation of a bond in the iPEPS we use the so-called full update (see \cite{corboz2010}) which is more accurate than the simple update~\cite{jiang2008,corboz2010} used in our previous study of the $t$-$J$ model~\cite{corboz2011}.

\begin{figure}[]
\begin{center}
\includegraphics[width=1\columnwidth]{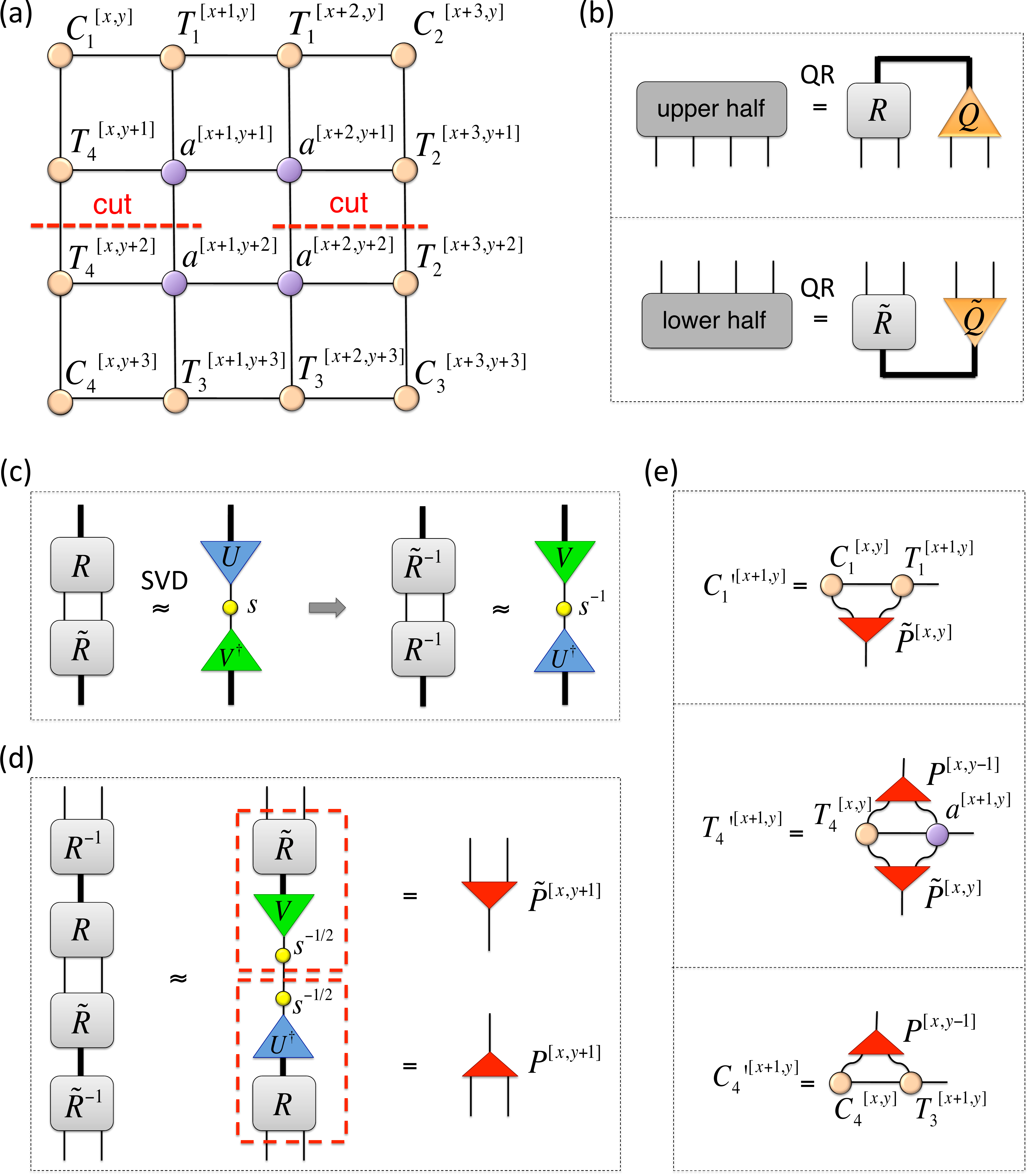} 
\caption{Details on the renormalization step in the corner-transfer-matrix method to perform a left move (cf. Ref.~\cite{corboz2011}). (a) A $2\times 2$ block of reduced tensors $a$ surrounded by the environment tensors. This network of $4\times4$ tensors is split into an upper half and a lower half, shown in (b). A QR decomposition (or SVD) is performed on the upper and lower half, yielding the tensors $R$ and $\tilde R$, respectively. (c) An SVD is performed on the product of $R \tilde R$, where only $\chi$ singular values are kept. (d)~A resolution of the identity $R^{-1} R \tilde R \tilde R^{-1}$ is approximated by introducing the result from (c), which yields projectors $\tilde P = \tilde R V s^{-1/2}$ and $P = s^{-1/2} U^\dagger R$. (e) These projectors are then used to absorb a column of tensors into the left environment tensors. 
}
\label{fig:ctm}
\end{center}
\end{figure}

\subsection{Contraction scheme}
In the present work we adopted the corner-transfer-matrix (CTM) method~\cite{nishino1996, orus2009-1}, generalized to arbitrary supercell sizes from Ref.~\cite{corboz2011}, to approximately contract the two-dimensional tensor network. The CTM method outputs the so-called environment tensors, consisting of four corner tensors $C_1$, $C_2$, $C_3$, $C_4$, and four edge tensors $T_1$, $T_2$, $T_3$, $T_4$, for each position $[x,y]$ in the supercell. These environment tensors effectively account for the infinite two-dimensional system surrounding the reduced bulk tensors $a^{[x,y]}$ (which are obtained by multiplying each tensor $A^{[x,y]}$ with its conjugate). For details on the method we refer to Ref.~\cite{corboz2011}. 

The only difference to the scheme in Ref.~\cite{corboz2011} is how we renormalize the corner and edge tensors after an absorption step. Instead of computing isometries based on a singular value decomposition to absorb a column (or a row) of tensors into the environment tensors, we use the projector introduced in Refs.~\onlinecite{Wang11, Huang12}. 
This choice of renormalization yields a better convergence of quantities as a function of the boundary dimension $\chi$.  The steps of how to perform a left-move (i.e. where the system is grown by one column to the left) is explained in Fig.~\ref{fig:ctm}. The other moves (right move, top move, and down move) are performed in a similar way until convergence is reached. 

A computationally cheaper variant of this (but less accurate) is to compute the QR decomposition shown in Fig.~\ref{fig:ctm}(b) only based on the upper left corner (made of 4 tensors), and lower left corner (made of 4 tensors), instead of the upper and lower half of the $4\times4$ network.

\subsection{Convergence of the variational energies}
\begin{figure}[]
\begin{center}
\includegraphics[width=1\columnwidth]{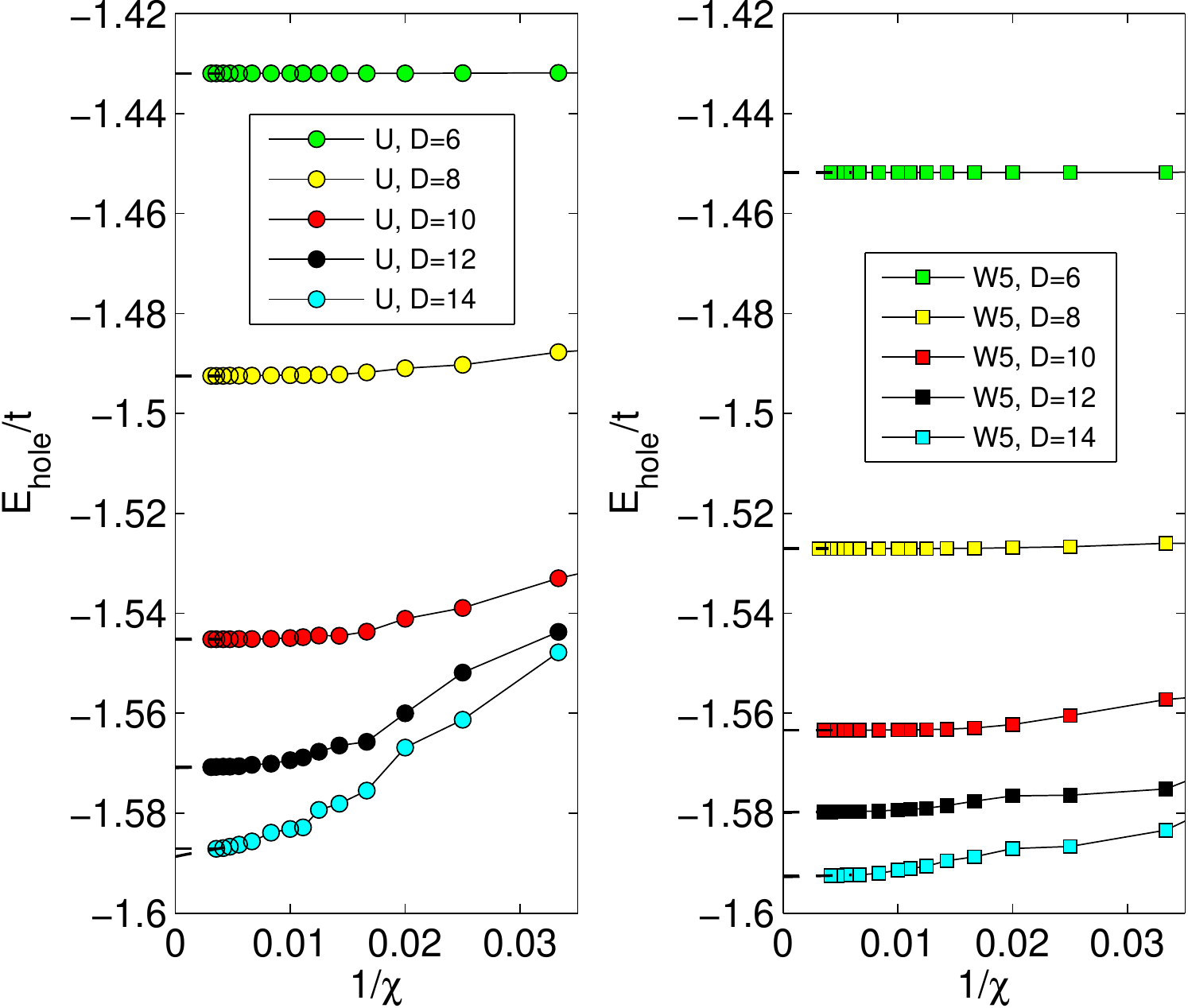} 
\caption{Convergence of the energies with the boundary dimension $\chi$ for different bond dimensions $D$ ($J/t=0.4$ and $\delta \sim 0.1$). The error due to the finite $\chi$ is small for large $\chi$.}
\label{fig:res_ntJchipaper}
\end{center}
\end{figure}

An iPEPS is a variational ansatz - however, expectation values can only be efficiently computed in an approximate way, with an error that is controlled by the boundary dimension $\chi$, see previous section. If $\chi$ is too small then the resulting energy is not necessarily variational, i.e. an upper bound to the true ground state energy. It is thus important to check the convergence of the energy (and order parameters) as a function of $\chi$, as e.g. shown in Fig.~\ref{fig:res_ntJchipaper}. In the present study we used a $\chi$ up to $\sim 300$ such that the error due to the finite $\chi$ is small (much smaller than the symbol sizes in the main text). Furthermore, in the present case we find that the energy is decreasing with increasing $\chi$, which implies that each energy $E(D,\chi)$ is variational (this is not necessarily true in general and needs to be  carefully checked  in each case).

\section{Additional simulation results}
\subsection{Isotropic case}
\begin{figure}[tb]
\begin{center}
\includegraphics[width=1\columnwidth]{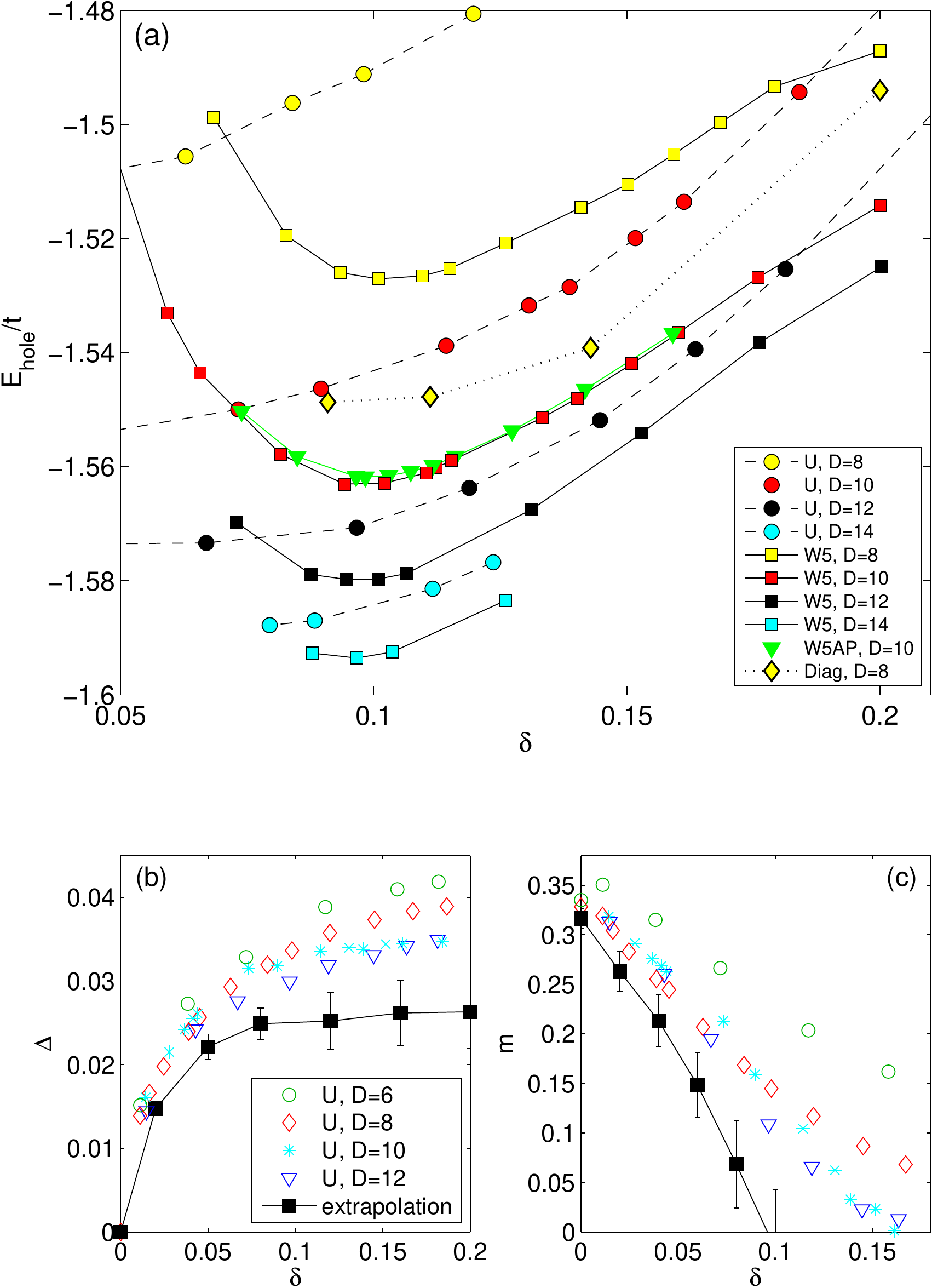} 
\caption{(a) Comparison of the variational energies for $J/t=0.4$ of the competing states for different bond dimensions $D$. (b) Pairing amplitude and (c)  magnetic moment in the uniform state as a function of doping. The extrapolated values have been obtained from a linear extrapolation of the finite $D$ data, which provides a rough estimate of the order parameters in the infinite $D$ limit.
}
\label{fig:res_ntJ04apapersuppl}
\end{center}
\end{figure}

In Fig.~\ref{fig:res_ntJ04apapersuppl}(a) we show the variational energies of several competing states as a function of doping for various values of $D$ for $J/t=0.4$. In particular, it shows that the anti-phase stripe W5AP has a similar or only slightly higher energy than the in-phase W5 stripe (for $D=10$). 
Figures~\ref{fig:res_ntJ04apapersuppl}(b)-(c) show the finite $D$ data of the magnetic moment and the pairing amplitude of the uniform state, together with the extrapolated value.

\begin{figure}[tb]
\begin{center}
\includegraphics[width=1\columnwidth]{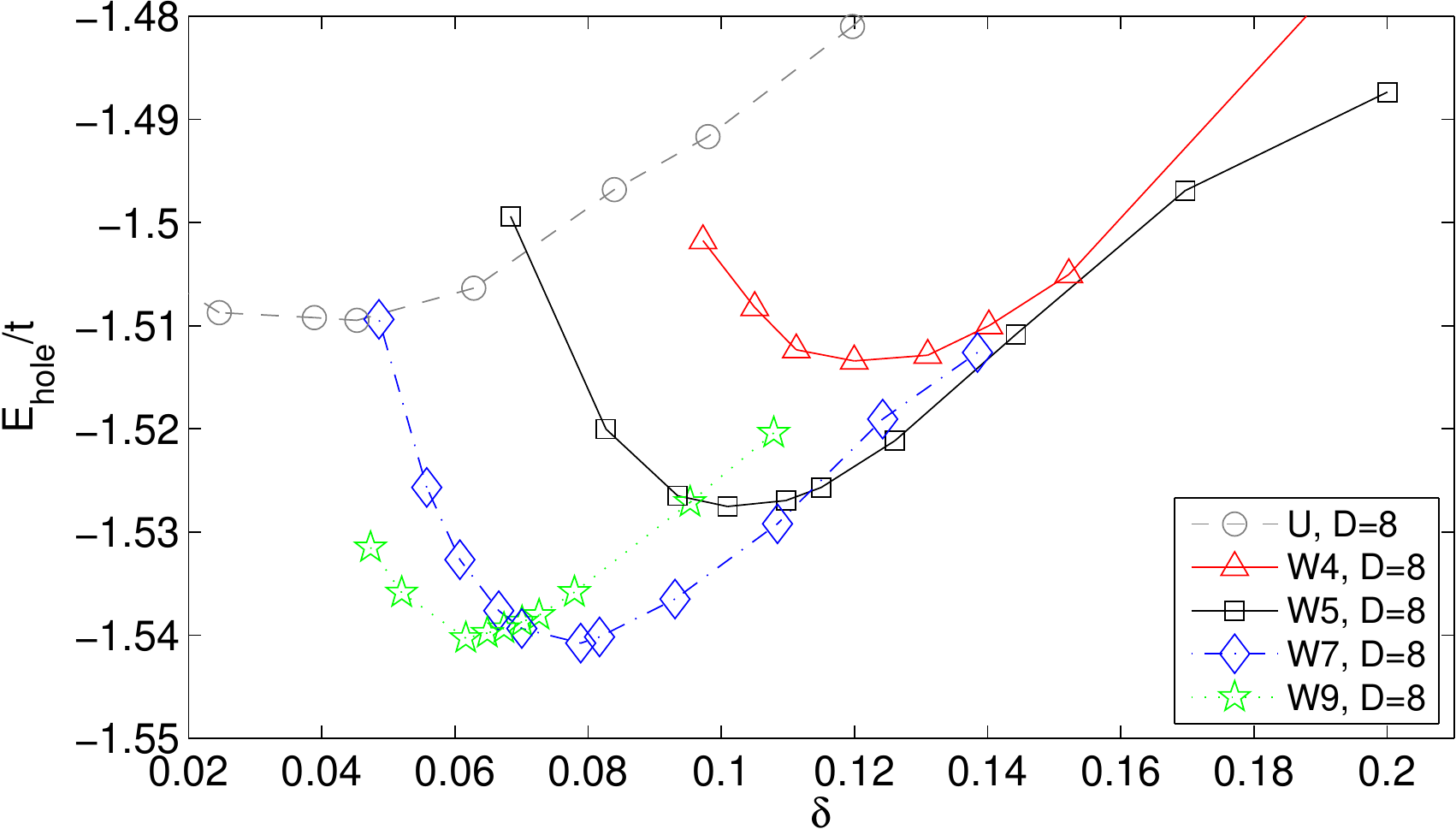} 
\caption{Energy per hole of various  stripes with widths $W$=4, 5, 7 and 9 (all in-phase and site centered) for $J/t=0.4$ and $D=8$.  }
\label{fig:res_ntJ04widepaper}
\end{center}
\end{figure}

In agreement with Ref.~\cite{corboz2011} we also find low-energy stripes in  supercells with other widths $W$, shown in Fig.~\ref{fig:res_ntJ04widepaper} for $D=8$. For each stripe the optimal filling (where the energy per hole is minimal) is  approximately at $\rho_l\approx 0.5$ holes per unit length per stripe for this particular value of $J/t=0.4$.  
The energy at the optimal stripe filling decreases with increasing stripe width up to $W=7$, but seems to saturate for larger widths, i.e. the minimal energy per hole of the $W=7$ stripe is roughly the same as the one of the $W=9$ stripe.  Note also that around $\delta\sim 1/8$ the $W=5$ stripe is lower in energy than the $W=4$ stripe. For dopings around $\delta\sim 0.15$ they are almost equal in energy. These findings may be different for other values of $J/t$ or in more realistic models including a finite $t'$.

\begin{figure}[tb]
\begin{center}
\includegraphics[width=1\columnwidth]{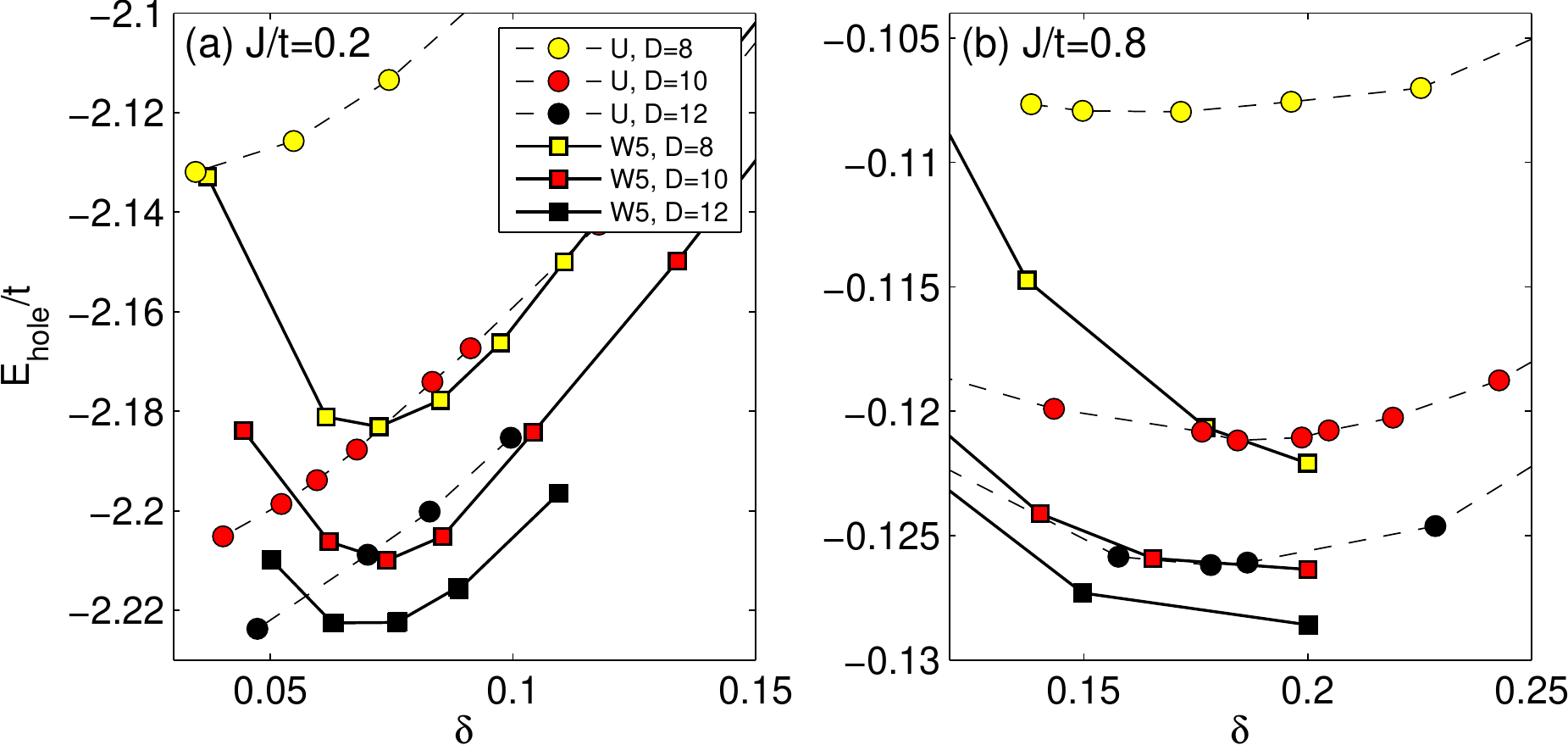} 
\caption{Variational energies of the W5 stripe and uniform state for (a) $J/t=0.2$ and (b) $J/t=0.8$. }
\label{fig:res_ntJ02paper3a}
\end{center}
\end{figure}

In Fig.~\ref{fig:res_ntJ02paper3a} we show the energies of the uniform and W5 stripe state for other values of $J/t$. The minimum energy per hole in the stripe state is found around $\rho_l\approx0.35$ ($\delta=0.07$) for $J/t=0.2$ and at $\rho_l=1$ ($\delta=0.2$) for~$J/t=0.8$. 

%
\subsection{Remarks on phase separation}
One of the possible predicted scenarios in the low-doping regime of the $t$-$J$ model is that the system phase separates into two phases, i.e. an antiferromagnetic region with $\delta=0$ and a doped region with a finite hole density (see e.g. Refs.~\cite{hellberg97, ivanov04} and references therein). For a stable system the energy per site must be a convex function as a function of doping, or equivalently the energy per hole needs to be a monotonically increasing with doping (see e.g.~Ref.~\cite{emery90}). If the energy per hole has a minimum at a certain value $\delta_c$ it implies that the system phase separates for hole densities  $0<\delta<\delta_c$. 

For example, if we consider the uniform state in Fig.~\ref{fig:res_ntJ04widepaper} we can identify a minimum in the energy per hole around $\delta_c\approx 0.045$ which would suggest an unstable region for $0<\delta<0.045$. This value changes with increasing $D$, e.g. $\delta_c \approx 0.03$ for $D=10$, and it is conceivable that $\delta_c$ tends to zero in the infinite $D$ limit.

However, the global minimum for $D=8$ in Fig.~\ref{fig:res_ntJ04widepaper} is given by the W7 stripe state for $\delta_c \approx 0.08$, which would suggest phase separation for $\delta\lesssim 0.08$, but not for $\delta>0.08$. Note that this global minimum may change with increasing $D$, e.g. the uniform state or stripes with larger widths (e.g. the W9 stripe) could become lower than the W7 stripe state for larger $D$, such that $\delta_c$ shifts to lower doping. A detailed study of the low doping regime is more challenging since it requires simulations using even larger supercell sizes and is left for future work.

For our study, the most important conclusion is that there is no evidence for phase separation for $\delta\gtrsim 0.08$ (for $J/t=0.4$), in particular not around $\delta=0.12$ at which we compare the energies of the competing states in the main text.

\begin{figure}[tb]
\begin{center}
\includegraphics[width=1\columnwidth]{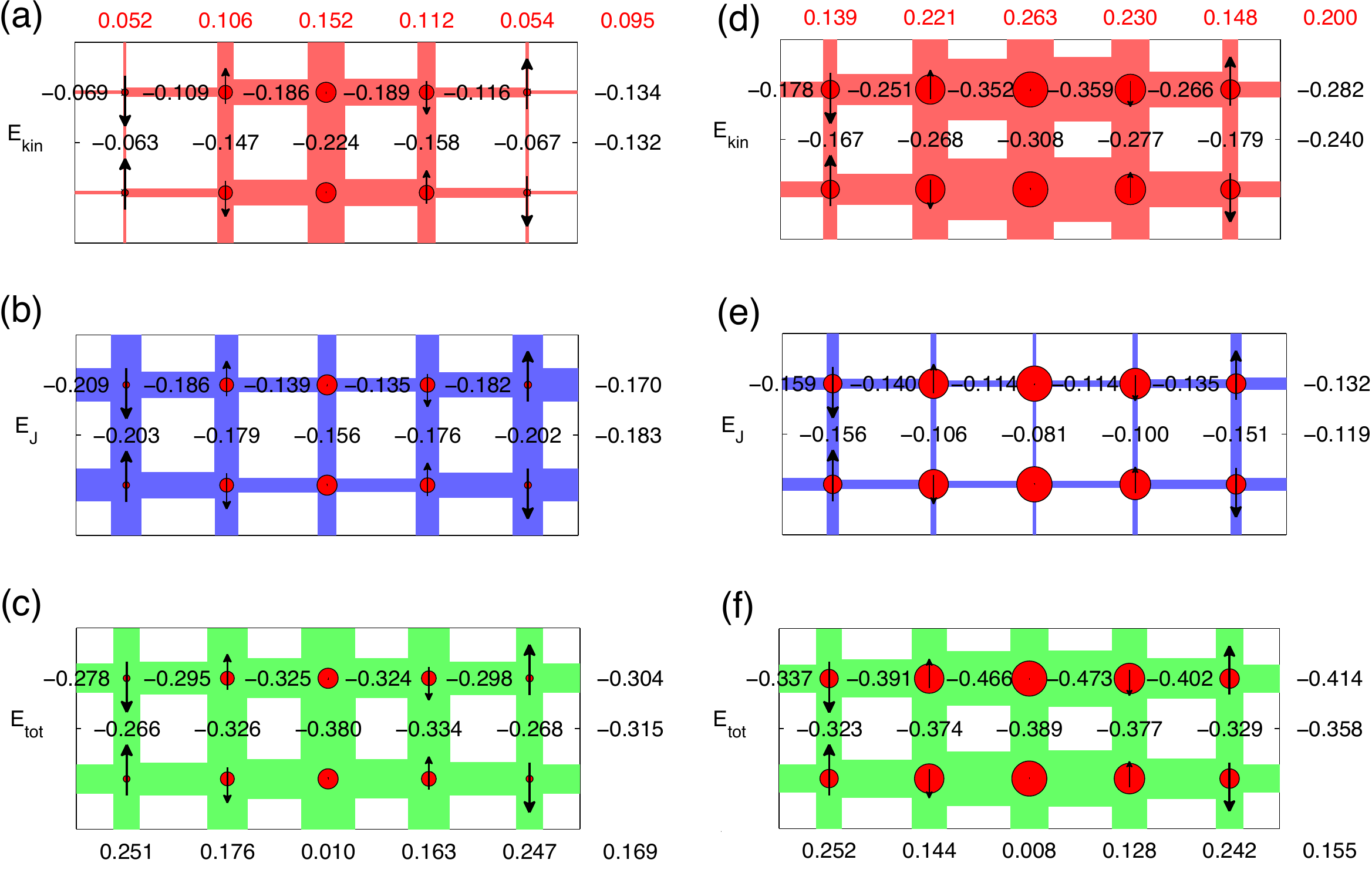} 
\caption{Energy contributions on each bond in the W5 stripe at doping $\delta=0.095$ (a)-(c), and $\delta=0.2$ (d)-(f). The first row of panels shows the kinetic energy $E^{kin}$, the second row the exchange energy $E^{J}$, and the last row the total energy one each bond in the supercell. The top (bottom) row of numbers show the local hole density (local magnetic moment). Mean values are given on the right hand side of each panel.  }
\label{fig:Econtr}
\end{center}
\end{figure}

\subsection{Energy contributions in the W5 stripe state}
In Fig.~\ref{fig:Econtr} we show the individual bond energies in the supercell of the W5 stripe at two different dopings, for $J/t=0.4$. As mentioned in the main text, at small doping ($\delta\sim0.1$) the most dominant energy is the exchange energy parallel to the stripe, $E^J_y$. Interestingly, the weakest exchange energy is not found along the vertical chain with highest doping, but along the x-direction on the bonds which connect to the vertical chain with highest doping. This indicates that along that chain rather strong AF correlations are present. 
At large doping ($\delta\sim 0.2$) the transverse kinetic energy $E^{kin}_x$ is dominant, and the exchange energy is also stronger along the $x$ than the $y$-direction. The weakest exchange energy is now found along the vertical chain with highest doping, and $E^J_x<E^J_y$, leading to a total energy which is clearly lower in $x$-direction.

\begin{figure}[tb]
\begin{center}
\includegraphics[width=1\columnwidth]{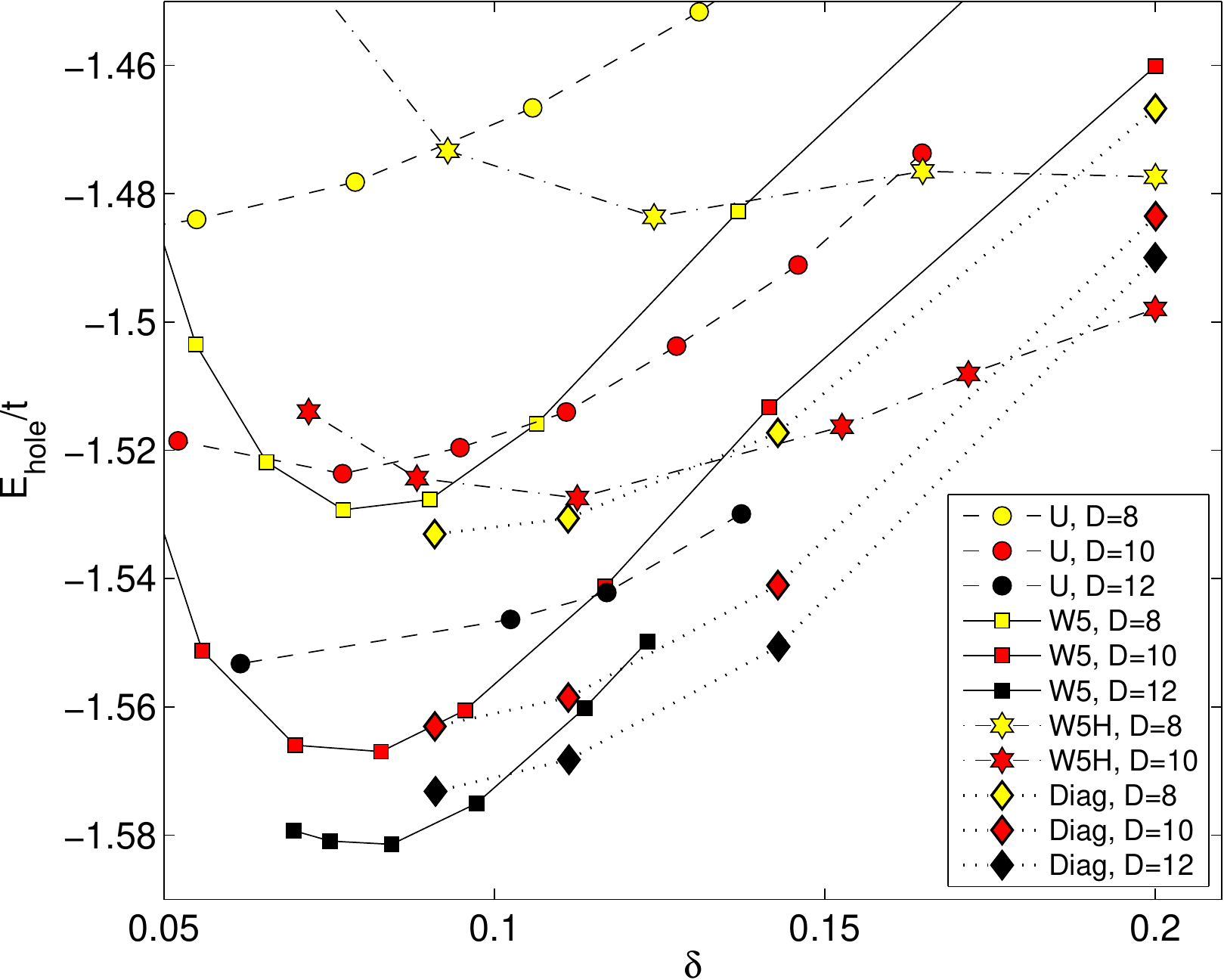} 
\caption{Comparison of the variational energies in the nematic $t$-$J$ model with $J/t=0.4$ and $t_x=0.85 t$ for different values of $D$. The W5H stripe corresponds to a width-5 stripe oriented along the horizontal direction (i.e. a W5 stripe rotated by 90 degrees).}
\label{fig:res_ntJ04tapapersupp}
\end{center}
\end{figure}

\subsection{Nematic case}
In Fig.~\ref{fig:res_ntJ04tapapersupp} we show the variational energies of the competing states for different values of $D$ ($J/t=0.4$, $t_x=0.85t$, and $t_y=t$). As in the isotropic case the insulating diagonal stripe has the lowest variational energy for small $D$, but the vertical stripe state becomes energetically lower for large $D$ around optimal filling. We did not push the simulations to large $D$ around $\delta\sim0.14$, but also in this case we expect that the superconducting horizontal or vertical stripe will be lower in energy than the insulating diagonal stripe, i.e. there is a transition between vertical and horizontal stripe as a function of doping.

\bibliographystyle{apsrev4-1}
\bibliography{../bib/refs}